\documentclass[
twocolumn,
 amsmath,amssymb,
 aps, physrev,
]{revtex4-2}
\usepackage{orcidlink}
\usepackage[svgnames]{xcolor} 
\definecolor{MyLightBlue}{RGB}{0,0,255}

\usepackage{graphicx}
\usepackage{dcolumn}
\usepackage{bm}

\begin{document}

\preprint{APS/123-QED}

\title{\textbf{Two-body cluster entangled structure in the $\mathcal{H}_1\otimes \mathcal{H}_2$ Hilbert space} 
}%

\author{Fei-Long Xu\orcidlink{0009-0009-8188-5480}}
\affiliation{Shanghai Institute of Applied Physics, Chinese Academy of Sciences, Shanghai 201800, China.}
\affiliation{University of Chinese Academy of Sciences, Beijing 101408, China.}
\affiliation{Shanghai Advanced Research Institute, Chinese Academy of Sciences, Shanghai 201210, China.}

\author{Xi-Guang Cao\orcidlink{0000-0003-2795-9601}}
\email{caoxg@sari.ac.cn}
\affiliation{Shanghai Advanced Research Institute, Chinese Academy of Sciences, Shanghai 201210, China.}
\affiliation{Shanghai Institute of Applied Physics, Chinese Academy of Sciences, Shanghai 201800, China.}
\affiliation{University of Chinese Academy of Sciences, Beijing 101408, China.}

\author{Yu-Gang Ma\orcidlink{0000-0002-0233-9900}}
\email{mayugang@fudan.edu.cn}
\affiliation{Key Laboratory of Nuclear Physics and Ion-beam Application (MOE), Institute of Modern Physics, Fudan University, Shanghai 200433, China.}
\affiliation{Shanghai Research Center for Theoretical Nuclear Physics, NSFC and Fudan University, Shanghai 200438, China.}
\affiliation{Shanghai Institute of Applied Physics, Chinese Academy of Sciences, Shanghai 201800, China.}
\affiliation{University of Chinese Academy of Sciences, Beijing 101408, China.}

\date{\today}

\begin{abstract}

This study introduces a quantum information perspective to analyze the internal structure of atomic nuclei, focusing on the quantum entanglement between $\alpha$ clusters in the 0$^+$ state of $^8$Be.
A wave function based on angular momentum coupling is developed to transform the two-cluster wave function from the conventional center of mass and relative coordinate basis ($\mathcal{H}_{R_{c.m.}} \otimes \mathcal{H}_r$) into the individual particle basis ($\mathcal{H}_1 \otimes \mathcal{H}_2$), which is essential for a precise quantification of entanglement.
Within this method, the von Neumann entropy is employed to quantify the entanglement arising from the mixing of angular momentum channels. Additionally, we introduce the concept of spatially‐resolved entropy, which measures entanglement as a function of the radial separation between clusters. 
Our analysis reveals that the entanglement is strongly correlated with the spatial configuration at the femtometer scale. As the inter cluster separation vanishes, the system approaches a separable $S$-wave state, indicating that entanglement is dynamically generated during the spatial separation of the clusters. This research provides a new tool for investigating nonlocal quantum correlations in nuclear structure, complementing existing descriptions.

\end{abstract}

\maketitle


\section{\label{sec:level1}Introduction}
In the 1930s, Wefelmeier, along with Wheeler and others, pioneered the study of molecular-type structures within atomic nuclei \cite{wefelmeier1937geometrisches,wheeler1937molecular}. Among these efforts, the resonating group method (RGM) developed by Wheeler became the theoretical foundation for various nuclear cluster models \cite{wheeler1937mathematical}, such as the generator coordinate method (GCM) and the orthogonality condition model. In these models, the $\alpha$-cluster structure has been particularly widely studied \cite{he2014giant,cao2019examination}.

In recent years, the Tohsaki-Horiuchi-Schuck-Röpke (THSR) method, an extension of the Brink wave function concept, has successfully described exotic cluster structures in nuclei such as $^{12}$C, $^{16}$O, and ${^{20}}$Ne \cite{tohsaki2001alpha,zhou2013nonlocalized,von2006nuclear}. However, these exotic cluster configurations, like the famous Hoyle state of $^{12}$C, are fundamentally many-body quantum systems. This essential characteristic implies that quantum entanglement, a key feature of many-body quantum systems, should play a crucial role in their structure \cite{itoh2014further,dell2017high,smith2017new,otsuka2022alpha}.

In theoretical nuclear physics, approximating the wave function as separable is a powerful and widely used simplification.
While the separable wave function approximation has been highly effective, it may not fully address the complexities of many-body nonlocality.
The physical reality of quantum entanglement, and its inherent nonlocality, is now empirically indisputable. A series of landmark experiments, typified by the loophole-free Bell's inequality test by Hensen \textit{et al.}, has decisively refuted local realism and established the nonlocality of entangled states as a fundamental principle of nature \cite{hensen2015loophole}. This experimentally confirmed nonlocality offers a powerful new perspective for analyzing nuclear systems.

Introducing the quantum information into nuclear physics holds the promise of providing new insights into our understanding of the atomic nucleus and may resolve some controversies \cite{ioffe1999precision,dell2017high}.This means we can understand clusters not only as geometric arrangements but also through the quantum correlations that bind them. Therefore, this paper aims to introduce this new perspective by re-examining the issue of locality among ideal $\alpha$ bosons and conducting a quantitative study of entanglement in the 0$^+$ ground state of $^8$Be.

\section{Theoretical formalism}

In the theoretical description of two-body quantum systems, the selection of an appropriate basis representation is crucial for simplifying the problem and elucidating specific physical insights. In quantum information science, the quantitative analysis of a system's entanglement properties is typically defined within the tensor product space of the constituent subsystems' Hilbert spaces, $\mathcal{H}_1 \otimes \mathcal{H}_2$ \cite{nielsen2010quantum}. This representation is naturally suited for discussing the system's separability and nonlocal correlations.

On the other hand, in nuclear physics, for systems dominated by an interaction potential of the form $V(|\mathbf{r}_1 - \mathbf{r}_2|)$, employing the Hilbert space corresponding to the center of mass $(\mathbf{R}_{c.m.})$ and relative coordinates $(\mathbf{r})$, denoted as $\mathcal{H}_{R_{c.m.}} \otimes \mathcal{H}_{r}$, is more convenient. This approach allows for the separation of the Hamiltonian into two parts, the free motion of the center of mass and the internal relative motion, thereby simplifying the solution process.

Nevertheless, applying the concept of quantum entanglement, which is naturally defined in the $\mathcal{H}_1 \otimes \mathcal{H}_2$ basis, to nuclear physics systems presents significant practical challenges. There remain considerable difficulties in efficiently performing the basis transformation for complex nuclear wave functions obtained in the $\mathcal{H}_{R_{c.m.}} \otimes \mathcal{H}_{r}$ representation and subsequently calculating their entanglement measures under a specific subsystem partition.
Therefore, the core task of this research is to construct a suitable basis vector to achieve an effective transformation between these two critical representations.

\subsection{Wave functions based on $\mathcal{H}_{1} \otimes \mathcal{H}_{2}$}
The two-body cluster theory presented in this paper is built upon the RGM, a highly precise cluster model first proposed by Wheeler in 1937 \cite{wheeler1937mathematical}. Within this framework, the wave function for a system composed of two clusters, $C_1$ and $C_2$, is expressed as,
\begin{equation}
\label{eq1}
\begin{aligned}
&\Phi\left(C_1+C_2\right)=\phi_{c.m.}\left(\boldsymbol{R}_{c.m.}\right)\mathcal{A}\left\{\chi(\boldsymbol{r})  \phi_1{(C_1)}\phi_2{(C_2)}\right\},\\
      &\boldsymbol{r}=\boldsymbol{r}_{1}-\boldsymbol{r}_{2},
\end{aligned}
\end{equation}
where $\boldsymbol{r}_{1} $and $\boldsymbol{r}_{2}$ are c.m. coordinates of clusters $C_1$ and $C_2$, respectively.
A$_k$ stands for the mass number of cluster $C_k$ and A represents the total mass number. 
$\mathcal{A}$ is the antisymmetrization operator, $\chi\left(r\right)$ is the relative wave function, $\phi_1 ({ C_{1}})$ and $\phi _2({ C_{2}})$  are the intrinsic wave functions of cluster $C_1$ and cluster $C_2$.
Our investigation is not concerned with the intrinsic state of nuclear matter within each cluster, but is instead directed towards the properties of the relationship between the clusters. The wave function,
\begin{equation}
\label{eq2}
 \psi\left( \mathbf{r};\mathbf{R}_{c.m.} \right) =\phi_{c.m.}\left(\boldsymbol{R}_{\mathrm{c.m.}}\right)u(r)Y_{J,M},
\end{equation}
 is therefore described within the Hilbert space constructed as $\mathcal{H}_{R_{c.m.}} \otimes \mathcal{H}_{r}$. ${u_{l_r}}(r)$ is the wave function for the antisymmetrization between clusters.
 
The total angular momentum of the system can be described in the spaces $\mathcal{H}_{\mathbf{R}_{c.m.}} \otimes \mathcal{H}_{\mathbf{r}}$ and $\mathcal{H}_1 \otimes \mathcal{H}_2$ as
\begin{equation}
\label{eq3}
   \boldsymbol{ \hat J} = \boldsymbol{{\hat J_{{R_{{\rm{c.m.}}}}}}} + \boldsymbol{{\hat J_r}} = \boldsymbol{{\hat J_1}} + \boldsymbol{{\hat J_2}},
\end{equation}
where $\boldsymbol{\hat{J}}$ is the total angular momentum, $\boldsymbol{\hat {J}_{R_{c.m.}}}$ is the center-of-mass angular momentum, $\boldsymbol{\hat{J}_r}$ is the relative angular momentum, and $\{\boldsymbol{\hat{J_i}}\}$ is the angular momentum of $\{C_i\}$ in the $\mathcal{H}_1 \otimes \mathcal{H}_2$ representation.
For a system with zero intrinsic angular momentum (spin), the total angular momentum $\boldsymbol{J}$ is simply the orbital angular momentum $\boldsymbol{l}$.
The condition (\ref{eq3}) guarantees also the conservation of parity in the wave function as
\begin{equation}
    \label{eq4}
    ( - 1)^{l_r + l_{R_{c.m.}}}=( - 1)^{{l_1} + {l_2}} .
\end{equation}
In the center-of-mass frame, we can obtain ${ J_{{R_{{\rm{c.m.}}}}}} = 0$. The correspondence between the Hamiltonian and the angular momentum can be described as 
\begin{equation}
    \label{h}
    \left[ {H,\boldsymbol{\hat J}} \right] = \left[ {H,{\boldsymbol{{\hat L}_r}}} \right] = \left[ {H,{\boldsymbol{{\hat L}_1}} + {\boldsymbol{{\hat L}_2}}} \right] = 0.
\end{equation}
This indicates that the energy eigenstates are eigenstates of the system's  total angular momentum $\boldsymbol{\hat J}$. Each cluster under consideration is described by a discrete orbital angular momentum. The energy eigenstates described in the Hilbert space $\mathcal{H}_1 \otimes \mathcal{H}_2$ are still eigenstates of the total angular momentum formed by the coupling of angular momenta $l_1$ and $l_2$. Therein, the wave function of the angular momentum eigenstates of $\boldsymbol{\hat{L}}_1+\boldsymbol{\hat{L}}_2$  for the 0$^+$ state of $^8$Be can be described as,
\begin{equation}
    \mathcal{Y}_{\{\alpha,J=0M=0\}}:={\left[ {{Y_{\alpha ,-m}} \otimes {Y_{\alpha ,m}}} \right]_{\{\alpha, 00\} }},
\end{equation}
where $\{\alpha ,JM\}$ represent the different coupling channels (indexed by $\alpha$) for two particles coupling to form the total angular momentum ($J$,$M$).
For the 0$^+$ state of $^8$Be, the specific form of the coupled channels can be described as
\begin{equation}
\label{eq7}
  \begin{aligned}
&\left[ {{Y_{l_1, m_1}} \otimes {Y_{l_2,m_2}}} \right]_{\{ \alpha ,00\} } ={\left[ {{Y_{\alpha, -m}} \otimes {Y_{\alpha ,m}}} \right]_{\{ \alpha ,00\} }}\\
&=\sum_{m=-\alpha}^\alpha\left(\begin{array}{ccc}
\alpha & \alpha & 0 \\
-m & m & 0
\end{array}\right) Y_{\alpha, -m}\left(\hat{r}_1\right) Y_{\alpha,m}\left(\hat{r}_2\right).
    \end{aligned}
\end{equation}
Therefore, the two-body cluster wave function represented in the $\mathcal{H}_{1} \otimes \mathcal{H}_{2}$ Hilbert space can be described as
\begin{equation}
\label{eq8}
    \begin{aligned}
\psi &= \phi_{c.m.}(R_{c.m.})  u_{l_r}(r) Y_{00}\\
&= {\sum\limits_\alpha  {{\Phi _\alpha }\left( {{r_1},{r_2}} \right)\left[ {{Y_{\alpha,-m}}\left( {{{\hat r}_1}} \right) \otimes {Y_{\alpha,m}}\left( {{{\hat r}_2}} \right)} \right]} _{\{ \alpha ,00\} }},
\end{aligned}
\end{equation}
where $\Phi_\alpha(r_1,r_2)$ is the radial wave function of the cluster. The states $Y_{\alpha, -m}\left(\hat{r}_1\right)Y_{\alpha,m}\left(\hat{r}_2\right)$ form an orthonormal basis of the state spaces $\mathcal{H}_1\otimes \mathcal{H}_2$.
This form, represented by Eq. (8) as $|\psi\rangle_{A B}=\sum_{i, j} c_{i j}|i\rangle_A \otimes|j\rangle_B$, is precisely the form of quantum entanglement \cite{horodecki2009quantum}. 
In the following section, we will provide a detailed demonstration to show that the transformation is rigorous near the origin.  This will show  that the coupled-channel form in Eq. (\ref{eq8}) can be obtained not only from the eigenstates of angular momentum but also directly from a mathematical transformation.

\subsection{Analysis of local transformation properties of wave function }
In the RGM, the antisymmetrized relative wave function can be obtained from the eigenvalues of the norm kernel \cite{horiuchi1977chapter, saito1977chapter}. This process can be described as
 \begin{align}
    {\chi _{l_r}}(r) &= \sum\limits_n {{a_n}} {R_{n,{l_r}}}({b_r};r), \\
    {u_{l_r}}(r) &= \sum\limits_n {{a_n}} \sqrt {{\mu _{n,{l_r}}}} {R_{n,l_r}}({b_r};r),
    \end{align}
where $a_n = \langle \chi_{l_r}(r) \mid R_{n,l_r}(b_r; r) \rangle$
and $R_{n,l_r}(b_{r};r)$ are the radial wave functions of the harmonic oscillator with the width parameter $b_r=1/\sqrt{2\gamma_0}=\sqrt{(A_1+A_2)/A_1A_2}b$. ${u_{l_r}}(r)$ is the antisymmetrized wave function.
$\mu_N$ depends on $N(2n+l_r)$ which is the eigen value of the RGM norm kernel. That can be described as
\begin{equation}
\label{eq11}
     \mu_N=\left\langle X_{N}(\boldsymbol{r}, \gamma_0) \phi_1 (C_1)\phi_2(C_2)\mid \mathcal{A}\left\{X_{N}(\boldsymbol{r}, \gamma_0) \phi_1 (C_1)\phi_2(C_2)\right\}\right\rangle,
\end{equation} 
where $X_N(\boldsymbol{r},\gamma_0)$ is the normalized three-dimensional harmonic oscillator function of oscillator constant $\gamma_0$ with N quanta only in the Z direction, namely,
\begin{equation}
\label{eq12}
X_N(\boldsymbol{r}, \gamma_0)=B_N H_N\left(\sqrt{2 \gamma_0} Z_r\right) e^{-\gamma_0 r^2}, \\
\end{equation}
\begin{equation}
\label{eq13}
    B_N=\left(\frac{2 \gamma_0}{\pi}\right)^{3 / 4}\left(2^N N!\right)^{-1 / 2}.
\end{equation}
$H_N(z)$ is the Hermite polynomial. The relationship between $R_{n,l_r}(b_{r};r)$ and $X_N(\boldsymbol{r}, \gamma_0)$ is established through angular momentum projections. That can be described as
\begin{equation}
\label{eq14}
    R_{n,l_r}(b_r, r) Y_{l_r 0}\left(\Omega_r\right)\propto \int d \Omega {D_{00}^{l_r}}^*(\Omega) \hat{R}(\Omega) X_N(\boldsymbol{r}, \gamma_0).
\end{equation}
For the $0^+$ state of $^8$Be, its $R_{n,l_r=0}$ can be described as
\begin{equation}
\label{eq15}
    {R_{n,0}} = {b^{-3/2}}{\sqrt 2 {{( - 1)}^n}{e^{ - \frac{{{r^2}}}{{2{b^2}}}}}\sqrt {\frac{{n!}}{{\Gamma \left( {n + \frac{3}{2}} \right)}}} L_n^{\frac{1}{2}}\left( {\frac{{{r^2}}}{{{b^2}}}} \right)},
\end{equation}
where $L_n^{\frac{1}{2}}\left(\frac{r^2}{b^2}\right)$ is the associated Laguerre polynomial and $\Gamma\left(n+\frac{3}{2}\right)$ is the gamma function \cite{kalzhigitov2023resonance}.
For the 0$^+$ state of $^8$Be, GCM and THSR used the same $b=1.36$ fm parameter \cite{funaki2002description}.
We perform a Taylor expansion of it with respect to $r^2$ around the origin.
This process enables us to examine the structure of the wave function from a more localized viewpoint.
Its expanded form can be described as
\begin{equation}
\label{eq16}
    {R_{n,0}} = \sum\limits_{p = 0} {{T_{p,n}}{r^{2p}}} ,
\end{equation}
where 
$T_{p, n}=\left.\frac{1}{( p)!} \frac{d^{p} R_{n, 0}}{d (r^{2}) {^p}}\right|_{r=0}$.
When we perform a convergence test on the series, it is straightforward to determine that $R_n(r)$ is an entire function. Its Taylor series expansion at any point, including $r=0$, will converge over the entire complex plane. Therefore, its radius of convergence is infinite .
By applying the binomial theorem to $(r^2)^p$, the power form of the relative distance vector can be expanded as a series sum. This form allows for the separation of the radial coordinates $r_1,r_2$ from the azimuthal angles $\Omega_1,\Omega_2$. That can be described as,
\begin{equation}
\label{eq17}
\begin{aligned}
    {r^{2p}} &=\left( r_1^2 + r_2^2 - 2{r_1}{r_2}\cos \gamma \right)^p
\\&= \sum\limits_{k = 0}^p {\left( \begin{array}{c}
p\\
k
\end{array} \right)} \left( {r_1^2 + } \right.{\left. {r_2^2} \right)^{p - k}}{\left( { - 2{r_1}{r_2}} \right)^k}{(\cos \gamma )^k},
\end{aligned}
\end{equation}
where $\gamma$ is the angle between vectors $\mathbf{r}_1$ and $\mathbf{r}_2$. Then we perform an orthogonal expansion of $(\cos \gamma )^k$ in terms of Legendre polynomials ${P_l}(\cos \gamma )$, and this process can be described as
\begin{equation}
\label{eq18}
(\cos \gamma )^k = \sum\limits_{l = 0, k - l \text{= even}}^k {\frac{{k!(2l + 1)}}{{(k - l)!!(k + l + 1)!!}}} {P_l}(\cos \gamma ).
\end{equation}
The relationship between the spherical harmonics of two directions and the Legendre polynomial of the angle between them is established via the multipole expansion, which is common in electromagnetism.
\begin{equation}
\label{eq19}
\begin{aligned}
    &{P_l}(\cos \gamma ) \\
 &= \frac{{4\pi }}{{2l + 1}}\sum\limits_{m =  - l}^l {{{\left( { - 1} \right)}^m}{Y_{l, - m}}} ({\theta _1},{\phi _1}){Y_{l,m}}({\theta _2},{\phi _2})\\
 &= \frac{{4\pi }}{{2l + 1}}{\left( { - 1} \right)^{ - l}}\sqrt {2l + 1} {\left[ {{Y_{l, - m}}(\hat r_1) \otimes {Y_{l,m}}(\hat r_2)} \right]_{\{ l,00\} }}
\end{aligned}
\end{equation}
Therefore, the relative wave function $u_{r}$ can be characterized by a multiple series expansion in the form of 
\begin{equation}
\label{eq20}
\begin{aligned}
&{u_l}(r) = \sum\limits_n {{a_n}} \sqrt {{\mu _{n,0}}} {R_{n,0}}({b_r};r)\\
&= \sum\limits_n {{a_n}} \sqrt {{\mu _{n0}}} \left( {\sum\limits_{p = 0} {{T_{p,n}}{r^{2p}}} } \right)\\
& = \sum\limits_{n = 0} {\sum\limits_{p = 0} {\sum\limits_{k = 0}^p {\sum_{\substack{l = 0 \\ k - l \text{= even}}}^k {{f _{n,p,k,l}}\left( {{r_1},{r_2}} \right){{\left[ {{Y_{l, - m}}(\hat r_1) \otimes {Y_{lm}}(\hat r_2)} \right]}_{\{ l,00\} }}} } } } \\
&= \sum\limits_l{{\Phi_{r,l}}\left( {{r_1},{r_2}} \right){{\left[ {{Y_{l, - m}}(\hat r_1) \otimes {Y_{l ,m}}(\hat r_2)} \right]}_{\{ l,00\} }}},
\end{aligned}
\end{equation}
where the general term of the series is 
\begin{equation}
\label{eq21}
\begin{aligned}
&f _{n,p,k,l}(r_1, r_2) \\
&= a_n \sqrt{\mu_{n,0}} T_{p,n} 
\left(\begin{array}{c}
p\\
k
\end{array}
\right)
\left( r_1^2 + r_2^2 \right)^{p - k} \left( -2r_1r_2 \right)^k \\
&\quad \times \frac{4\pi (-1)^{-l} \sqrt{2l + 1} k!}{(k - l)!! (k + l + 1)!!}.
\end{aligned}
\end{equation}
The form of ${\Phi _{l_r}}\left( {{r_1},{r_2}} \right)$ is determined by the multiple summation in Eq. (\ref{eq20}).
Within the GCM, the center-of-mass wave function can be described as
\begin{equation}
\label{eq22}
    \phi_{c.m.}=\left(\frac{(A_1+A_2) \gamma_0}{\pi}\right)^{3 / 4} \exp \left\{-\frac{1}{2} \gamma_0(A_1+A_2) \boldsymbol{R}_{\mathrm{c.m.}}^2\right\}.
\end{equation}
Using a similar strategy, the centroid wave function can also be represented as
\begin{equation}
\label{eq23}
    \phi _{c.m.} = \sum_{\beta} \Phi _{R,\beta}(r_1, r_2) \left[ Y_{\beta, -m}(\hat{r}_1) \otimes Y_{\beta, m}(\hat{r}_2) \right]_{\{\beta, 00\}}
\end{equation}
Thus, the wave function for the two-body cluster within the framework of the single-particle Hilbert space can be rewritten as
\begin{equation}
\label{eq24}
    \begin{aligned}
    \psi&=\phi_{c.m.}\left(\boldsymbol{R}_{c.m.} \right){u_{l_r}}\left( r \right){Y_{00}}\\
     &\propto\sum\limits_{\beta} {{\Phi _{R,\beta}}\left( {{r_1},{r_2}} \right){{\left[ {{Y_{{\beta }, - m}}(\hat r_1) \otimes {Y_{{\beta},m}}(\hat r_2)} \right]}_{\{ {\beta},00\} }}}\\
     &\times \sum\limits_l {{\Phi _{r,l} }\left( {{r_1},{r_2}} \right){{\left[ {{Y_{l , - m}}(\hat r_1) \otimes {Y_{l,m}}(\hat r_2)} \right]}_{\{ l,00\} }}}\\
     &=\sum\limits_\alpha {{\Phi _\alpha }\left( {{r_1},{r_2}} \right){{\left[ {{Y_{\alpha, - m}}(\hat r_1) \otimes {Y_{\alpha, m}}(\hat r_2)} \right]}_{\{ \alpha,00\} }}}.
    \end{aligned}
\end{equation}
A more detailed proof of Eq. (\ref{eq24}) can be found in Appendix. Thus, through a local analysis at the origin, we consider this transformation in Eq. (\ref{eq8}) to be mathematically rigorous.

\subsection{Discussion on the physical interpretation of mathematical structures}
 Before numerically analyzing the transformation characterized by Eq. (8), we will first discuss its physical significance in detail here. The purpose of this is to demonstrate that this transformation process is not merely a brute-force mathematical construct.

 The description of a two-body quantum system within the framework of the Hilbert space  $\mathcal{H}_1 \otimes  \mathcal{H}_2$ has a long history.  For example, the well-known Talmi-Moshinsky brackets \cite{moshinsky1959transformation} effect the transformation for a harmonic oscillator potential between the basis of individual particle coordinates ($\mathcal{H}_1\otimes \mathcal{H}_2$) and the basis of center-of-mass and relative coordinates ($\mathcal{H}_{R_{c.m.}}\otimes \mathcal{H}_r$). 
The description of the wavefunction in the Hilbert space $\mathcal{H}_1\otimes\mathcal{H}_2$ is established upon the discretization of single-particle angular momentum and the formalism of angular momentum coupling. This transformation process can be denoted by
\begin{equation}
\label{eq25}
\begin{aligned}
|n_{1}&l_{1},n_{2}l_{2},JM\rangle\\
&=\sum_{n_rl_rn_Rl_R}|n_rl_r,n_Rl_R,JM\rangle\langle n_rl_r,n_Rl_R,J|n_{1}l_{1},n_{2}l_{2},J\rangle
\end{aligned}
\end{equation}
and
\begin{equation}
\label{eq26}
\begin{aligned}
|n_r&l_r,n_Rl_R,JM\rangle\\
&=\sum_{n_1l_1n_2l_2}|n_{1}l_{1},n_{2}l_{2},JM\rangle\langle n_{1}l_{1},n_{2}l_{2},J| n_rl_r,n_Rl_R,J\rangle.
\end{aligned}
\end{equation}
$\{n_i\}$ represents the vibrational quantum number of the harmonic oscillator. $n_r$ and $n_R$ are the vibrational quantum numbers of the relative wave function and the center-of-mass wave function, respectively.
When the interaction potential between nucleons cannot be approximated as a harmonic oscillator, the $\{n_i\}$ quantum numbers in the Talmi-Moshinsky transformation may no longer correspond to a clear physical quantity.
However, for a system with conserved total angular momentum, we can also describe its wavefunction in the coupled basis. The structure of these basis vectors is independent of the interaction potential.

Shifting the perspective from relative and center-of-mass coordinates to those of individual particles does not signify a mechanical approach. Eq. (\ref{eq8}) is formulated upon the foundational principles of the quantization of orbital angular momentum and the equivalence between these two coordinate bases.
$\phi_{c.m.}(R_{c.m.}) u(r) Y_{00} $ and $ {\sum\limits_\alpha{{\Phi _\alpha }\left( {{r_1},{r_2}} \right)\left[ {{Y_{\alpha,-m}}\left( {{{\hat r}_1}} \right) \otimes {Y_{\alpha,m}}\left( {{{\hat r}_2}} \right)} \right]} _{\{ \alpha ,00\} }} $
respectively provide us with two perspectives to examine nuclear structure. 

\section{The inseparable property of the intercluster wave function}

Von Neumann explored the density matrix for quantum measurements, leading to the development of von Neumann entropy as an extension of Gibbs entropy for measuring the uncertainty of quantum systems \cite{von2013mathematische}. Notably, von Neumann entropy is remarkably similar in its mathematical form to Shannon entropy, which was later developed independently by Claude Shannon in information theory. Both of these entropy-based analytical methods have been widely applied in diverse fields such as nuclear physics \cite{shannon1948mathematical,ma1999application}.
 In this work, von Neumann entropy is used to measure the entanglement of two-body particles. For a pure state $\rho_{AB} = |\psi\rangle\langle\psi|_{AB}$, the entanglement entropy can be described as
\begin{equation}
\label{eq27}
    \mathcal{S}\left(\rho_A\right)=-\operatorname{Tr}\left[\rho_A \ln \rho_A\right]=-\operatorname{Tr}\left[\rho_B \ln \rho_B\right]=\mathcal{S}\left(\rho_B\right),
\end{equation}
where $\rho_A$ and $\rho_B$ are defined as $ \operatorname{Tr}_B(|\psi\rangle\langle\psi|_{AB})$ and $\operatorname{Tr}_A(|\psi\rangle\langle\psi|_{AB}) $, respectively.
\begin{figure}
\includegraphics[width=1\linewidth]{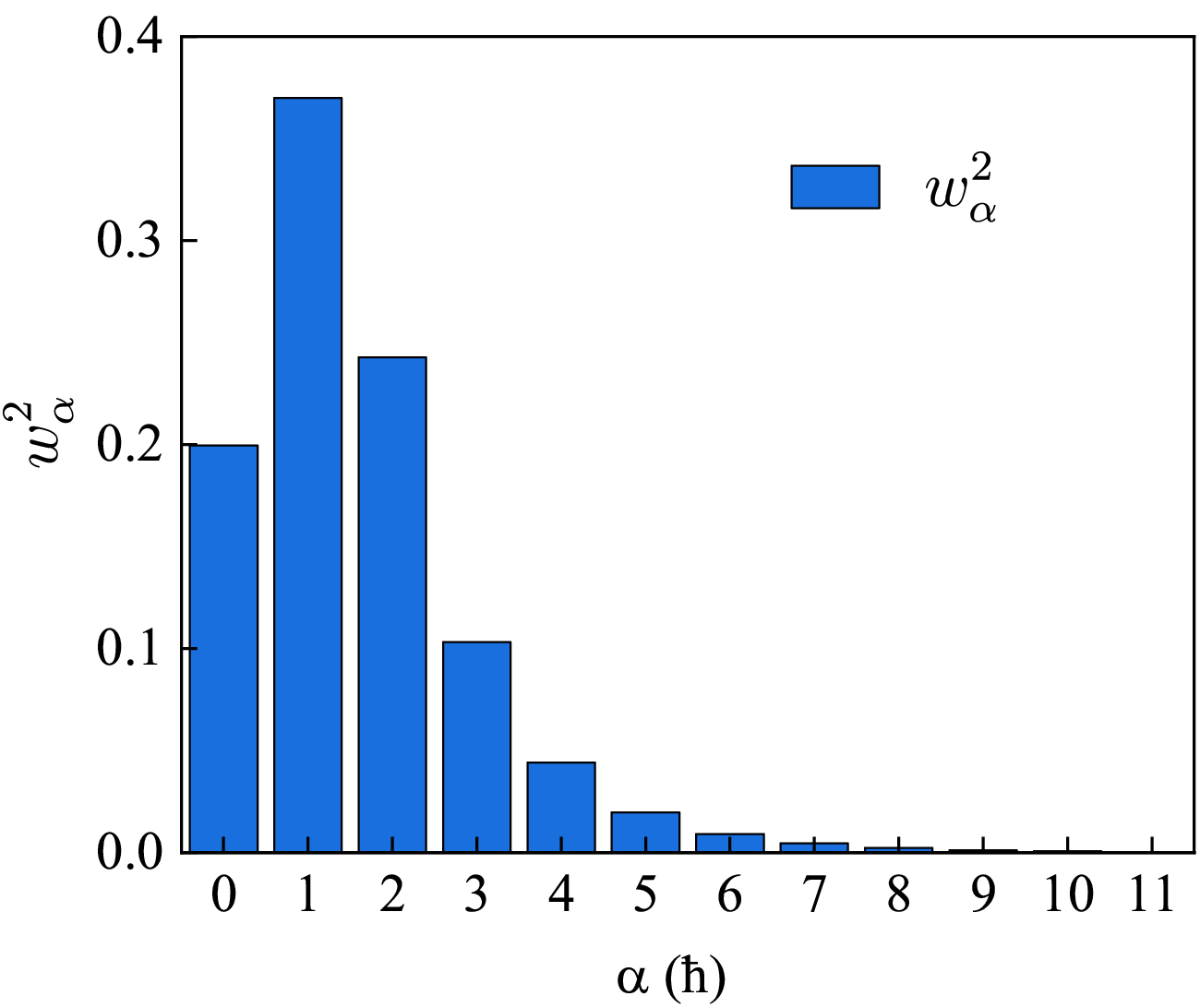}%
\caption{\label{fig0} 
The proportion of the $0^{+}$ coupled channels in $^{8}$Be}
\end{figure}

We use this entanglement entropy to measure the inseparability of the wave function described by Eq.  (\ref{eq8})  in the coupled-channel basis.
Thus, for the $0^+$ state of $^8$Be, the probability of each coupled channel $\alpha $ can be represented as
\begin{equation}
\label{eq28}
    {w^2_\alpha } = \int_0^\infty  {\int_0^\infty  {{{\left| {{\Phi _\alpha }\left( {{r_1},{r_2}} \right)} \right|}^2}} } r_1^2d{r_1}r_2^2d{r_2},
\end{equation}
where $\sum_{\alpha=0} ^\infty  w_\alpha^2=1$.
Therefore, the $\psi(\theta_1, \phi_1, \theta_2, \phi_2)$ in the tensor product $\mathcal{H}_1 \otimes \mathcal{H}_2$ can be uniquely determined by the Schmidt decomposition in terms of the orthonormal sets $\{Y_{lm}(\hat{r}_1)...\} \subset \mathcal{H}_1$ and $\{Y_{lm}(\hat{r}_2)...\} \subset \mathcal{H}_2$.
That can be described as
\begin{equation}
\label{eq29}
    \begin{aligned}
        \psi & = \sum_i\lambda _i u  _{i}\otimes v  _{i}\\
& = \sum_\alpha w_\alpha\left [ Y_{\alpha,-m} \otimes Y_{\alpha,m} \right ]_{\{\alpha,00 \}}\\
&=\sum_\alpha \sum_{m=-\alpha} ^{\alpha} w_\alpha \frac{(-1)^{\alpha-m}}{\sqrt{2\alpha +1}} Y_{\alpha,-m} \otimes Y_{\alpha,m},
    \end{aligned}
\end{equation}
where $\{\left| \lambda_i\right|\}$ are the Schmidt coefficients.
Using the relationship between the Schmidt coefficients and entanglement entropy, the entanglement generated by the mixing of different angular momentum channels $\alpha$ can be described as
\begin{equation}
\label{eq30}
\begin{aligned}
S\left( \rho_A \right) &=
 -Tr_{A}\left( \rho_A ln \rho_A \right)=-\sum_i \lambda^2_i \ln \left(\lambda^2 _i\right)\\
&=-\sum_\alpha \sum_{m=-\alpha}^\alpha\left(\frac{w_\alpha^2}{2 \alpha+1} \ln \frac{w_\alpha^2}{2 \alpha+1}\right)\\
&=-\sum_{\alpha=0} w_\alpha ^2\ln\left(\frac{{w_\alpha ^2}}{2\alpha +1}\right).
\end{aligned}
\end{equation}

This entropy quantitatively describes the strength of the quantum entanglement between the two $\alpha$ clusters within the $^8$Be nucleus, arising from their orbital motion. A higher entropy value signifies that the contributions of the various channels (with probabilities $w^2_\alpha$) are more uniformly distributed, making the superposition nature of the system more pronounced. Consequently, the quantum entanglement between the two $\alpha$ particles is stronger. The states of the two $\alpha$ clusters are thus more tightly correlated and cannot be described independently.

Eq. (\ref{eq24}) indicates that the total wave function can be rigorously expressed in a coupled-channels expansion. The radial wave function, $\Phi_\alpha(r_1,r_2)$, which can be obtained by projecting the total wave function onto the corresponding coupled channels, is given by
\begin{equation}
\label{eq31}
\begin{aligned}
\Phi_{\alpha'}(r'_1,r'_2) &= \left\langle \psi(\mathbf{r}_1, \mathbf{r}_2) \middle| \frac{\delta(r_1 - r'_1)\,\delta(r_2 - r'_2)}{{r'}_1^{2}\,{r'}_2^{2}}\right. \\
& \left. \times  \middle| [Y_{\alpha',-m'}(\hat{r}_1) \otimes Y_{\alpha',m'}(\hat{r}_2)]_{\alpha',00} \right\rangle.
\end{aligned}
\end{equation}
Thus, $w^2_\alpha$ can be obtained from Eq. (\ref{eq28}).
The contributions of the first 12 coupled channels to the center-of-mass wave function of the configuration in Eq. (\ref{eq22}) are shown in Fig. 1. The total probability accounted for by these channels is 0.996. Consequently, using Eq. (30), the orbital angular momentum entanglement for the $0^+$ state of the $\alpha$-$\alpha$ system is calculated to be 2.766.

\begin{figure}
\includegraphics[width=1\linewidth]{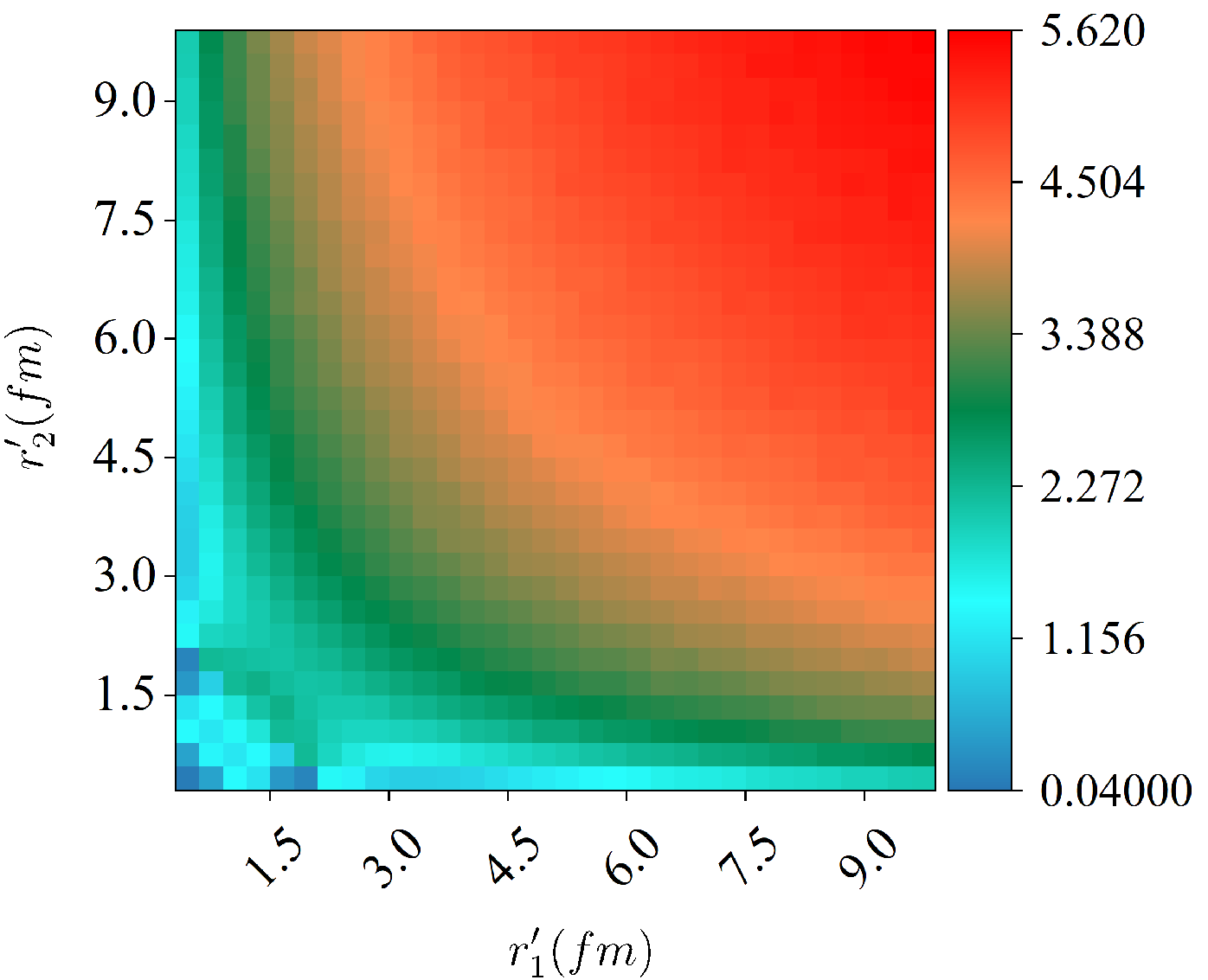}%
\caption{\label{fig1} 
Spatially-resolved entropy on different spatial radii }
\end{figure}

To complement the entropy that provides a holistic description of entanglement in $^8$Be, we introduce a spatially resolved entropy of angular momentum coupled channels at fixed radii to quantify its more localized properties. In the center-of-mass frame for a single-particle representation, the momentum is characterized by $\boldsymbol{p}_1 + \boldsymbol{p}_2 = 0$.
Simultaneously, the uncertainty principle leads to uncertainty in position, which changes the classical identity $\boldsymbol{r}_1 +\boldsymbol{r} _2 = 0$ into $\left\langle \boldsymbol{\hat{r}_1} + \boldsymbol{\hat{r}_2} \right\rangle = 0$.
This is evaluated when the clusters are at specific radial distances, $r_1'$ and $r_2'$, from their common center of mass. This approach provides a method to probe the nature of entanglement among nuclear constituents in a specific spatial snapshot.
To implement this entanglement measure, the wave function is first normalized at the given radii. The normalization factor, $F^2(r_1', r_2')$, is defined as
\begin{equation}
\label{32}
F^2(r_1', r_2') = \iint dV_1 dV_2  \psi^* \psi \frac{ \delta(r_1 - r_1') \delta(r_2 - r_2')}{r_1^2r_2^2} .
\end{equation}
Physically, $F^2(r_1',r_2')$ also represents the probability of finding the two $\alpha$ particles at the specified radii $r_1'$ and $r_2'$. At this radial configuration, the probability amplitude for the system to be in a specific angular momentum coupled channel, denoted by $\alpha$, is given by 
\begin{equation}
\label{eq33}
\begin{aligned}
    \mathcal{W}_\alpha(r_1', r_2') &= \iint dV_1 dV_2 \frac{\psi}{F(r_1', r_2')} \left[ Y_{\alpha,- m} \otimes Y_{\alpha , m} \right]_{\{\alpha,00\}} \\
&\times \frac{ \delta(r_1 - r_1') \delta(r_2 - r_2')}{r_1^2r_2^2}.
\end{aligned}
\end{equation}
This is computed by projecting the radially normalized wavefunction onto the basis of coupled spherical harmonics.
Consequently, the entropy at the specified radii can be described as
\begin{equation}
\label{eq34}
S(r_1', r_2') = - \sum_\alpha \mathcal{W}_\alpha^2(r_1', r_2') \ln\left(\frac{\mathcal{W}_\alpha^2(r_1', r_2')}{2\alpha + 1}\right).
\end{equation}
This calculation is analogous to the standard von Neumann entropy but utilizes the probabilities ($\mathcal{W}_\alpha^2$) of the system occupying different angular momentum coupled channels at a fixed spatial separation. A higher value of $S(r_1',r_2')$ signifies  stronger and more complex channels coupling between the two $\alpha$-particles at these particular distances.

Figure 2 shows the spatially resolved entropy, $S(r_1', r_2')$, at different spherical radii, $r_1'$ and $r_2'$. The entropy exhibits a strong correlation with the fm scale. This indicates that from a dynamical perspective, the angular momentum entanglement of the two $\alpha$ particles is strongly correlated at the fm scale. The local oscillations in entropy are primarily concentrated within 3 fm, while at larger scales, the entropy shows a monotonically increasing trend.
\begin{figure}
\includegraphics[width=1\linewidth]{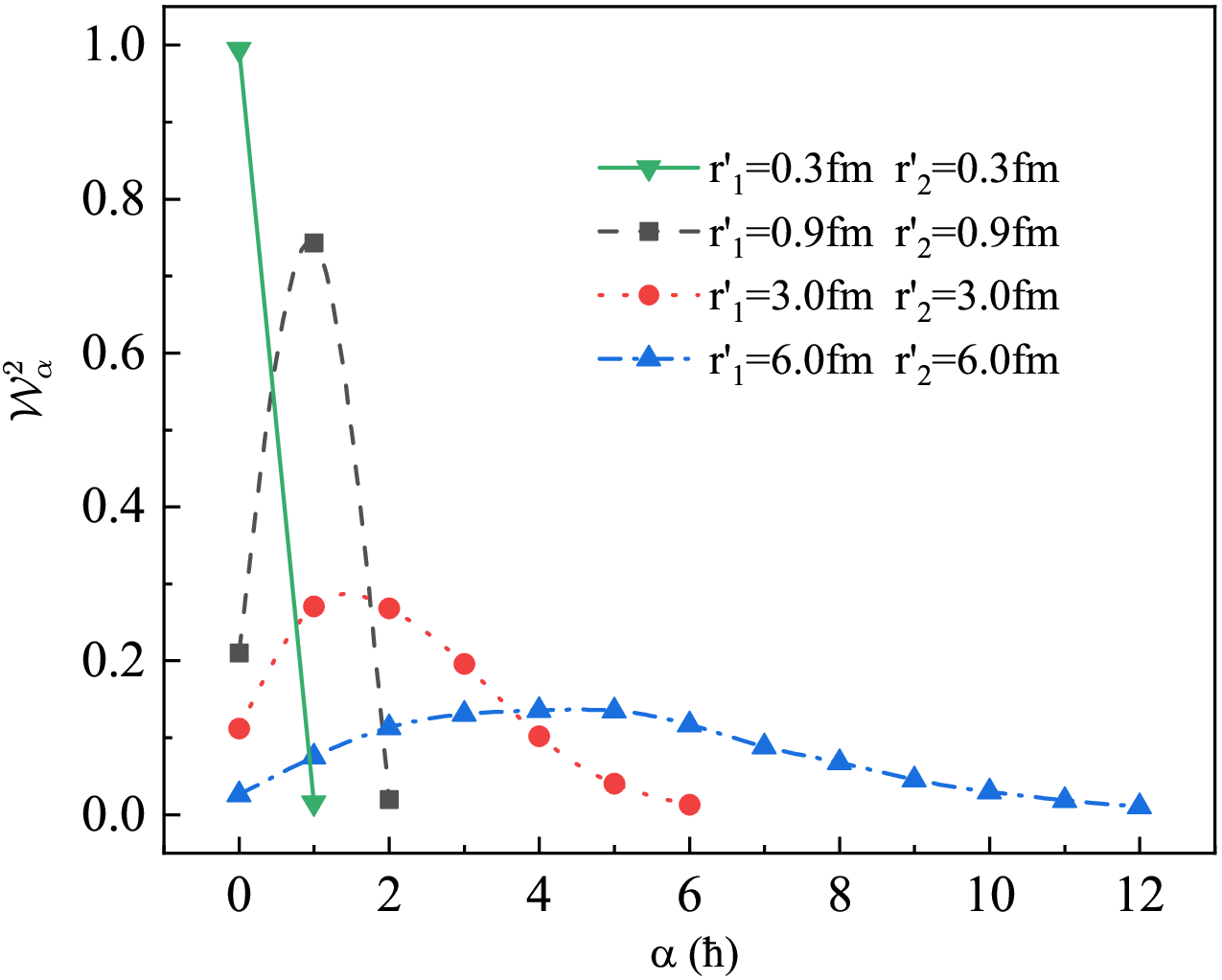}%
\caption{\label{fig2} 
The proportion of coupling channels at different spatial radii }
\end{figure}

Fig. 3 displays the proportional contributions of different coupled channels for $r_1$ and $r_2$ at various radii (in fm), covering a cumulative probability of  $\sum_\alpha \mathcal{W}_\alpha^2\gtrsim 0.99$.
It is evident that the number of contributing channels is limited and primarily concentrated in those with low angular momentum. As the radius increases, there is a discernible trend of contributions shifting towards coupled channels with higher angular momentum. This finite and small number of contributing channels, in turn, indicates that although we utilize a Taylor expansion and an orthogonal expansion of Legendre polynomials, these expansions are not merely mathematical approximations of an infinite series. 

Additionally, the asymptotic behaviors of $r_1$ and $r_2$ were analyzed. We found that when the distance between clusters approaches zero, the coupling of the wave function on the spherical surface gradually vanishes, and the \(0^+\) state of \(^{8}\mathrm{Be}\) can be described as $S_1\text{wave}\otimes S_2\text{wave}$. That can be described as
\begin{equation}
\label{eq35}
\begin{aligned}
&\lim_{r'_1 \to 0, r'_2 \to 0} \mathcal{W}_0(r'_1, r'_2) \\
&= \lim_{r'_{1}  \to 0, r'_2\to 0} \iint dV_{1}  dV_{2} \frac{\left(\frac{8 \gamma_0}{\pi}\right)^{3 / 4} \exp \left\{-4 \gamma_0 \boldsymbol{R}_{\mathrm{c.m.}}^2\right\}}{F(r_1', r_2')r_1^2r_2^2} \times  \\
&\sum\limits_n {{a_n}} \sqrt {{\mu _{n,{l_r}}}} {R_{n,l_r}}({b_r};r)\left[ Y_{00} \otimes Y_{00} \right]_{\{0,00\}}\delta(r_1 - r_1') \delta(r_2 - r_2')\\
&= 1.
\end{aligned}
\end{equation}
This suggests that the generation and evolution of spatially‐resolved entropy  are driven by the $S$ wave channel from a dynamical perspective.

Finally, we discuss the experimentally observable effects for ${\left[ {{Y_{\alpha ,-m}} \otimes {Y_{\alpha ,m}}} \right]_{\{\alpha, 00\} }}$. For a two-body wave function resulting from the decay of a resonance state, this form of angular momentum coupling is inherently present. This is a nonlocal phenomenon that is independent of the spatial radii of the two particles. Consequently, the asymptotic behavior of this two-body wave function at infinity in the center of mass frame is significantly different from the plane-wave description, ${e^{i\mathbf{ k}_c {\mathbf{r} _1}}} \otimes {e^{-i\mathbf{k}_c{\mathbf{r} _2}}}$, used for free particles. This difference may affect the subsequent quantum scattering process. That is to say, when each particle of the entangled pair bombards a new target nucleus as an incident particle, its original form of angular momentum coupling may be carried over and influence the new quantum scattering process.
Moreover, it is worth noting that Eq. (23) reveals that the entanglement entropy and the spatially resolved entropy are dependent on the configuration of the center-of-mass wavefunction.
The Brink-type wavefunction utilized herein describes a spatially confined system, such as the $^{8}\text{Be}$ nucleus produced in the decay of the $^{12}\text{C}$ Hoyle state.

\section{Summary and conclusions}
The debate over locality in quantum mechanics has a long history, yet entanglement in nuclear physics has been seldom studied. In this study, we construct a basis to investigate the characteristics of quantum entanglement within nuclear cluster structures, focusing on the 0$^+$ state of  $^8$Be. By bridging the gap between the traditional nuclear physics description in the center of mass and relative coordinate basis ($\mathcal{H}_{R_{c.m.}} \otimes \mathcal{H}_r$) and the quantum information framework based on the tensor product of individual cluster Hilbert spaces ($\mathcal{H}_1 \otimes \mathcal{H}_2$), we can quantitatively analyze the non-local features of the system.

Within this method, we compute the von Neumann entropy as a quantitative measure of entanglement between the two $\alpha$ clusters. An entanglement entropy $S$ quantifies the total entanglement in the $^8$Be ground state, while a newly introduced spatially resolved entropy, $S(r'_1, r'_2)$, captures the spatial dependence of entanglement. 

Our numerical results reveal several key insights based on the wave function of the Brink configuration. The $^8$Be ground state possesses significant quantum entanglement, arising from the superposition of multiple coupled angular momentum channels. This confirms that the two $\alpha$ clusters cannot be treated as independent, separable entities. The spatially resolved entropy analysis shows that entanglement is highly dependent on the distance between the clusters. It exhibits strong oscillatory behavior at short distances and grows monotonically at larger separations as higher angular momentum channels become more prominent. We demonstrated that as the inter-cluster distance approaches zero, the spatially resolved entropy vanishes, and the system is correctly described by two $S$ wave $\alpha$ particles.

In conclusion, this work establishes a robust and quantitative bridge between nuclear cluster theory and quantum information science. By recasting a classic nuclear physics problem in the language of entanglement, we have provided new, quantifiable insights into the nonlocal quantum correlations governing nuclear structure. Our framework offers a new tool for exploring the fundamental nature of the nucleus as a deeply entangled quantum many-body system.

\begin{acknowledgments}
This work was supported by the National Key Research and Development Program of China (Grant No. 2022YFA1602404), the Strategic Priority Research Program of Chinese Academy of Sciences (Grant No. XDB34030000), the National Natural Science Foundation of China (Grants No. 12475134, No. 12147101, No. 11975210, and No. U1832129
), and the Youth Innovation Promotion Association CAS (Grant No. 2017309).
\end{acknowledgments}

\appendix
\section{Product representation of the coupled-channel wave function}

This appendix derives the product representation of the coupled channel wave functions as presented in Eq.~(24) of the main text. The proof leverages the properties of spherical harmonics and the expansion of Legendre polynomials.

We begin with the coupled channel wave function \(\mathcal{Y}_{\{\alpha,00\}}\), which represents two individual angular momenta, each with quantum number \(\alpha\), coupled to a total angular momentum of zero. This is formally expressed using Clebsch-Gordan coefficients as
\begin{equation}
\begin{aligned}
\mathcal{Y}_{\{\alpha,00\}} &= \left[ Y_{\alpha,-m} \otimes Y_{\alpha,m} \right]_{\{\alpha,00\}}\\
&= \sum_{m=-\alpha}^{\alpha} \begin{pmatrix} \alpha & \alpha & 0 \\ -m & m & 0 \end{pmatrix} Y_{\alpha, -m}(\hat{
r}_1) Y_{\alpha, m}(\hat{r}_2).
\end{aligned}
\end{equation}
By applying the properties of Clebsch-Gordan coefficients and the conjugate relation of spherical harmonics, this expression simplifies to
\begin{equation}
\mathcal{Y}_{\{\alpha,00\}} = \frac{(-1)^{\alpha}}{\sqrt{2\alpha+1}} \sum_{m=-\alpha}^{\alpha} Y_{\alpha, m}^*(\hat{r}_1) Y_{\alpha, m}(\hat{r}_2).
\end{equation}
The summation term can be directly evaluated using the spherical harmonics addition theorem,
\begin{equation}
\sum_{m=-\alpha}^{\alpha} Y_{\alpha, m}^*(\hat{r}_1) Y_{\alpha, m}(\hat{r}_2) = \frac{2\alpha+1}{4\pi} P_\alpha(\cos \gamma),
\end{equation}
where \(P_\alpha\) is the Legendre polynomial, and \(\gamma\) is the angle between \(\hat{r}_1\) and \(\hat{r}_2\). Substituting this result yields a compact form for the coupled wave function,
\begin{equation}
\mathcal{Y}_{\{\alpha,00\}} = \frac{(-1)^{\alpha}}{\sqrt{2\alpha+1}} \cdot \frac{2\alpha+1}{4\pi} P_\alpha(\cos \gamma) = \mathcal{C}_\alpha
 P_\alpha(\cos \gamma),
\end{equation}
where the constant $\left\{\mathcal{C}_i\right\}$ is defined as $\left\{\frac{(-1)^{i} \sqrt{2i+1}}{4\pi}\right\}$.
Next, we examine the product of the two partial wave expansions that appear in Eq. (24) of the main text. The product is
\begin{equation}
\begin{aligned}
  & \left( \sum_{\beta} \Phi_{R,\beta}(r_1, r_2) \mathcal{Y}_{\{\beta,00\}} \right) \left( \sum_{l} \Phi_{r,l}(r_1, r_2) \mathcal{Y}_{\{l,00\}} \right) \\
    &= \sum_{\beta, l} \Phi_{R,\beta}(r_1, r_2) \Phi_{r,l}(r_1, r_2) \mathcal{Y}_{\{\beta,00\}} \mathcal{Y}_{\{l,00\}}.
\end{aligned}
\end{equation}
To simplify this expression, we must evaluate the product of the two coupled wave functions, \(\mathcal{Y}_{\{\beta,00\}} \mathcal{Y}_{\{l,00\}}\). We can achieve this by using the Legendre polynomial product expansion theorem,
\begin{equation}
P_\beta(x) P_l(x) = \sum_{\alpha=|\beta - l|}^{\beta + l} (2\alpha + 1) \begin{pmatrix} \beta & l & \alpha \\ 0 & 0 & 0 \end{pmatrix}^2 P_\alpha(x),
\end{equation}
Since \(\mathcal{Y}_{\{\beta,00\}} = \mathcal{C}_\beta P_\beta(\cos \gamma)\) and \(\mathcal{Y}_{\{l,00\}} = \mathcal{C}_l P_l(\cos \gamma)\), their product becomes
\begin{align}
\mathcal{Y}_{\{\beta,00\}} \mathcal{Y}_{\{l,00\}} &= \mathcal{C}_\beta \mathcal{C}_l P_\beta(\cos \gamma) P_l(\cos \gamma) \notag \\
&= \mathcal{C}_\beta \mathcal{C}_l \sum_{\alpha=|\beta - l|}^{\beta + l} (2\alpha + 1) \begin{pmatrix} \beta & l & \alpha \\ 0 & 0 & 0 \end{pmatrix}^2 P_\alpha(\cos \gamma).
\end{align}
Given that \(\mathcal{Y}_{\{\alpha
,00\}} = \mathcal{C}_\alpha P_\alpha(\cos \gamma)\), we can express the product as
\begin{equation}
\mathcal{Y}_{\{\beta,00\}} \mathcal{Y}_{\{l,00\}} = \sum_{\alpha=|\beta - l|}^{\beta + l} G(\beta, l, \alpha) \mathcal{Y}_{\{\alpha,00\}},
\end{equation}
where the coefficient \(G(\beta, l, \alpha)\) is defined as
\begin{equation}
G(\beta, l, \alpha) = \frac{\mathcal{C}_\beta \mathcal{C}_l}{\mathcal{C}_\alpha} (2\alpha + 1) \begin{pmatrix} \beta & l & \alpha \\ 0 & 0
 & 0
 \end{pmatrix}^2,
\end{equation}
representing the coupling strength between the angular momenta \(\beta\), \(l\), and \(\alpha\).

Substituting this into the original product, we have

\begin{equation}
\begin{aligned}
    &\sum_{\beta, l} \Phi_{R,\beta}(r_1, r_2) \Phi_{r,l}(r_1, r_2) \mathcal{Y}_{\{\beta,00\}} \mathcal{Y}_{\{l,00\}}\\
    &= \sum_{\beta, l} \Phi_{R,\beta}(r_1, r_2) \Phi_{r,l}(r_1, r_2) \sum_{\alpha} G(\beta, l, \alpha) \mathcal{Y}_{\{\alpha,00\}}.
\end{aligned}
\end{equation}
By exchanging the order of summation, we arrive at
\begin{equation}
\begin{aligned}
    &\sum_{\alpha} \left[ \sum_{\beta, l} \Phi_{R,\beta}(r_1, r_2) \Phi_{r,l}(r_1, r_2) G(\beta, l, \alpha) \right] \mathcal{Y}_{\{\alpha,00\}}\\
    &= \sum_{\alpha} \Phi_{\alpha}(r_1, r_2) \mathcal{Y}_{\{\alpha,00\}},
\end{aligned}
\end{equation}
where \(\Phi_{\alpha}(r_1, r_2) = \sum_{\beta, l} \Phi_{R,\beta}(r_1, r_2) \Phi_{r,l}(r_1, r_2) G(\beta, l, \alpha)\). This completes the derivation, demonstrating that the product of two partial wave expansions can be expressed as a linear combination of single angular wave functions, as stated in Eq. (24).


\bibliography{ref}

\end{document}